\documentclass[12pt]{article}
\usepackage{amsfonts}
\usepackage{graphicx}
\usepackage{epsfig}
\usepackage{amsmath}
\usepackage{amssymb}
\begin{document}
\begin{titlepage}
\begin{flushright}
DESY 11-229\\
SFB/CPP-11-72\\
LPN11-68
\end{flushright}

\begin{center}
{\Large {\bf 
{Neutral triple electroweak gauge boson production in the large extra-dimension model at the LHC
}}} 
\\[1cm]

M.\ C.\ Kumar$^{a,} $\footnote{kumar.meduri@desy.de} 
\hspace{.5cm} 
Prakash Mathews$^{b,} $\footnote{prakash.mathews@saha.ac.in}
\hspace{.5cm} 
V.\ Ravindran$^{c,} $\footnote{ravindra@hri.res.in}
\hspace{.5cm} 
Satyajit Seth$^{b,} $\footnote{satyajit.seth@saha.ac.in}
\\[1cm]

${}^a$ 
Deutsches Elektronen-Synchrotron DESY, Platanenallee 6, D-15738
Zeuthen, Germany
\\[.7cm]

${}^b$ Saha Institute of Nuclear Physics, 1/AF Bidhan Nagar, Kolkata 700 064, 
India
\\[.7cm]

${}^c$ Regional Centre for Accelerator-based Particle Physics\\ 
Harish-Chandra Research Institute, Chhatnag Road, Jhunsi,\\
Allahabad 211 019, India

\end{center}
\vspace{1cm}

\begin{abstract}
\noindent
We study the prospects of probing large extra dimension model at the LHC
through neutral triple gauge boson production processes. 
In theories with extra dimensions these processes result from the exchange 
of a tower of massive graviton modes between the SM particles.  We consider 
$\gamma\gamma\gamma$, $\gamma \gamma Z$, $\gamma Z Z$ and $ZZZ$ production 
processes, and present our results for various kinematic distributions at 
the LHC for $\sqrt{S}=14$ TeV.

\end{abstract}

\vspace{.7cm}

\end{titlepage}

\section{Introduction}

The results that would emerge from the Large Hadron Collider (LHC) are 
expected to reveal patterns under which the standard model (SM) is 
likely to be modified at the TeV scale.  At these high energies, LHC is
expected to unveil the mystery of large hierarchy between the electroweak 
and the Planck scales and also produce interesting new physics signals.  
Two well-studied scenarios of physics beyond the SM at the TeV scale are 
extra dimension theories and supersymmetry, which will be tested at the 
LHC among many others.  The large extra dimension model, proposed 
by Arkani-Hamed, Dimopoulos and Dvali (ADD) \cite{ADD}, circumvents the 
hierarchy problem by exploiting the geometry of extra spatial dimensions 
to find a mechanism to lower the Planck scale.  The tower of 
Kaluza-Klein (KK) modes as a result of gravity propagating the full 
$4+d$-dimensions would mediate scattering processes and hence lead to 
deviations from the SM predictions.
At colliders, exchange of virtual KK modes or emission of real KK
modes give rise to interesting phenomenological signals at the
TeV scale \cite{GRW,HLZ}.  Virtual effects of KK modes 
lead to the enhancement of the cross section of pair production in 
processes {\em viz.}\ di-lepton, di-gauge boson $(\gamma \gamma, 
~ZZ, ~W^+ W^-)$, dijet \cite{dilepton,diphoton,diphoton1,dizz,dijet} 
in some kinematic regions.   The real emission of 
KK modes leads to large 
missing $\not \!\! E_T$ signals {\em viz.} mono jet, mono photon, 
mono Z boson and mono $W^\pm$ boson \cite{mono}. 

The di-gauge boson final states have been extensively studied in the 
context of extra dimension models.  The triple gauge boson final 
state is an interesting new physics signal in some of the beyond
SM scenarios \cite{wlist}.  
In this paper we consider the neutral triple gauge boson production at the LHC and 
study how the ADD model would alter the SM expectation.  In the SM, the 
triple gauge boson final state is an important signal as it depends on the 
3-point and 4-point couplings among the gauge bosons which is a test of the 
electroweak theory.  This process in the SM has been studied to leading order (LO) 
\cite{golden,barger} and its extension to the next to leading order (NLO) 
was on the Les Houches 
wish list \cite{wlist} and has been finally achieved recently \cite{binoth, 
lazo, bozzi,bozzi1}.
The triple gauge boson production processes in the SM are 
the precise predictions of the electroweak gauge theory and gauge self-couplings. 
They are also potential background to many new physics models like
supersymmetry and Technicolor.  For example, $Z\gamma \gamma$ in SM is a 
background to signals with di-photons and missing transverse energy in  
gauge mediated super symmetric theories \cite{susy} and $\gamma\gamma\gamma$
production in SM is a background to one photon plus techni-pion \cite{tcolour}.
Processes with three gauge bosons can also come from the ADD model as gravitons
couple directly to gauge bosons of SM.
While mono-jet or di-lepton production is more sensitive to parameters of models with 
extra-dimensions compared to the triple gauge boson production, all these 
processes involve same universal coupling of gravity with the SM particles, 
and hence can provide equally important information about
the model.  Moreover, in discriminating physics beyond the SM namely SUSY 
or technicolour models using triple gauge boson production, 
one can not ignore the potential contributions resulting from models
with extra dimensions.

In this analysis we consider the process $P P\rightarrow V V V ~X$, where 
we restrict to the neutral gauge bosons $V=\gamma,Z$ and $X$ is some hadronic 
final state.  The following four final states are the subject of this analysis: 
(i) $\gamma \gamma \gamma$ (ii) $\gamma \gamma Z$ (iii) $\gamma Z Z$ and (iv) 
$Z Z Z$.  The case where $V=W^{\pm}$ will be part of a different paper 
\cite{tobe}.  

This paper is organised as follows: in section 2 we present the 
analytical calculation of the above mentioned processes with a brief 
introduction to the ADD model, section 3 is arranged to present the 
numerical results of our studies and finally we summarise the results 
in the last section.  
   
\section{Neutral triple gauge boson production}

The TeV scale colliders like the LHC are suitable for studying the
quantum gravity effects in the ADD model.
This model is defined in such a way that 
the SM particles are localised on a $3$-brane with negligible tension and 
only gravity can propagate in the full $4+d$ dimensions as it describes the 
geometry of the full space time.  The extra spatial dimensions $d$ can be 
compactified on a $d$-dimensional torus with a common radius $R/(2\pi)$ of 
macroscopic size.  As a consequence, the effective Planck scale ($M_P$) in 
4 space-time dimensions is related to the fundamental Planck scale ($M_S$) 
in $4+d$ dimensions as follows,
\begin{eqnarray}
 M_P^2 = C_d\ M_S^{2+d}\ R^d  ~~ ,
\end{eqnarray}
where $C_d = 2(4\pi)^{-\frac{d}{2}}/\Gamma(d/2)$. 
Although, the SM particles are localised on a $3$-brane, they can sense the 
extra dimensions by their interaction with gravity via the massive KK towers 
of the graviton. The corresponding interaction Lagrangian is the following,
\begin{eqnarray}
 \mathcal{L} = - \frac{\kappa}{2} \sum_{\overrightarrow{n}=0}^\infty 
T^{\mu\nu}(x)h^{\overrightarrow{n}}_{\mu\nu}(x)  ~~ ,
\label{lagrange}
\end{eqnarray}
where $\kappa = \sqrt{16\pi}/M_P$, $T^{\mu\nu}$ is the energy-momentum tensor 
of localised SM fields and $h^{\overrightarrow{n}}_{\mu\nu}$ denotes the 
massive KK graviton labeled by a $d$-dimensional vector $\overrightarrow{n}$ 
with all positive components. 
$h^{\overrightarrow{n}}_{\mu\nu}$ contains one spin-2 
state, $(n-1)$ spin-1 states and $n(n-1)/2$ spin-0 states,  
having mass,  
\begin{eqnarray}
 m_{\overrightarrow{n}}^2 = \frac{4\pi^2 \overrightarrow{n}^2}{R^2 } ~~ .
\end{eqnarray}
The zero mode of the KK tower corresponds to the massless graviton in the 
4 space-time dimensions. 
The effective graviton propagator, after summing over all KK states, can be 
expressed as,
\begin{eqnarray}
 {\mathcal{D}}_{eff}(s) &=& \sum_{\overrightarrow{n}}\frac{1}
{s-m_{\overrightarrow{n}}^2+i\varepsilon}  ~~ ,
\nonumber \\
 &=& \frac{1}{\kappa^2}\frac{8\pi}{M_S^4}\left(\frac{\sqrt{s}}
{M_S}\right)^{(d-2)}[\pi+2iI(\Lambda/\sqrt{s})] ~~ ,
\label{propagator}
\end{eqnarray}
where $s$ is the invariant mass of the boson pair resulting from the
decay of the KK mode and the function 
$I(\Lambda/\sqrt{s})$ is described in \cite{HLZ}, which depends on 
the ultra violet cut-off $\Lambda$.  
Although the interaction of KK modes with the SM 
particles is suppressed by the coupling $\kappa$ 
(Eq.\ (\ref{lagrange})), the cumulative effect of summing over large 
number of accessible KK modes (Eq.\ (\ref{propagator})) compensates 
the suppression, making the effective coupling significant enough to
have observable effects.
It is usual practice to set the UV cutoff $\Lambda=M_S$ and simplify 
the summation of virtual KK modes \cite{GRW,HLZ} to do the
phenomenology.  We follow the approach of \cite{HLZ} for this
analysis which retains the details of the number of extra dimensions. 

\begin{figure}[tbh]
\centerline{
\epsfig{file=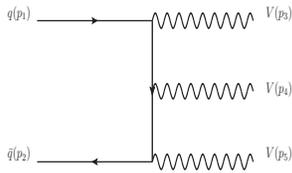,width=4cm,height=2.5cm,angle=0} }
\caption{Typical Feynman diagram for triple gauge boson production in SM.}
\label{lo_sm}
\end{figure}

\begin{figure}[tbh]
 \centerline{
\epsfig{file=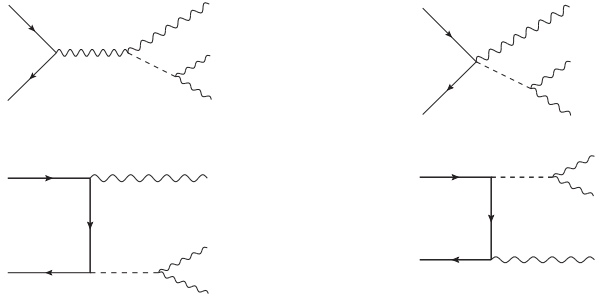,width=8cm,height=4cm,angle=0} }
\caption{Typical Feynman diagrams for triple gauge boson production in ADD 
model.  Dashed line represents the KK mode of the graviton and the other 
particle lines are same as in Fig.\ 1.}
\label{lo_bsm}
\end{figure}

The neutral gauge boson final state at the hadron collider 
$P P \to VVV ~X$ at LO comes from the sub process
\begin{eqnarray}
q(p_1) + \bar{q}(p_2) \longrightarrow V(p_3) + V(p_4) + V(p_5) ~~ ,
\end{eqnarray}
where $V = \gamma,Z$ and $X$ is any final state hadron.
The SM diagram for the above process is shown in Fig.\ \ref{lo_sm} with 
all possible permutations of final states.  For the final state with at 
least two $ZZ$s, 
the Higgs could contribute by coupling to the quarks, but this is negligible
in the vanishing quark mass limit.  In the case of $ZZZ$ final state, 
there are additional Higgs strahlung diagrams but their contribution 
is also quite small and vanishes in the large Higgs mass limit.  Hence, we have 
not included the processes with the Higgs boson.
In the ADD model, the KK modes couple to $V$ bosons, quarks, anti-quarks 
as well as to quark-antiquark-$V$ boson vertex. 
The four categories of Feynman diagrams that give a $VVV$
final state in ADD model are shown in Fig.\ \ref{lo_bsm}.
We have used unitary gauge ($\xi \rightarrow \infty$) for Z-boson and the
Feynman gauge ($\xi=1$) for the photon.

In the SM, the LO process for the production of $\gamma\gamma\gamma$
at hadron colliders results from the annihilation of a quark and an anti-quark. 
In the ADD model,
the production mechanism is again from the same initial states 
but one of the photon comes from either quark or an antiquark and other
two photons come from the decay of KK graviton.  
The typical Feynman diagrams that contribute in the SM and in the ADD model
are shown in Fig.\ \ref{lo_sm}, \ref{lo_bsm}.  
The Feynman rules for the processes with KK graviton 
can be found in \cite{GRW,HLZ}.  All the expressions for the matrix element 
squared with proper spin, color sums and averages
are obtained using a symbolic program based on FORM \cite{{FORM}}.  
The KK graviton propagator in the ADD model   
is proportional to ${\cal D}$,
\begin{eqnarray}
 {\cal{D}} &=& \sum_{\overrightarrow{n}}\frac{1}{s_{ij}-
m_{\overrightarrow{n}}^2+i\varepsilon}  ~~ ,
\label{prop1}
\end{eqnarray}
where the invariants are $s_{ij}=(p_i+p_j)^2$, and ${\cal{D}}$ can be 
evaluated using Eq.\ (\ref{propagator}).  The numerator of the spin-2 
propagator \cite{HLZ} of the KK graviton is given by 
\begin{eqnarray}
B_{\mu\nu,\rho\sigma}(k) &=& \eta_{\mu\rho} \eta_{\nu\sigma} 
+ \eta_{\mu\sigma} \eta_{\nu\rho} 
- \frac{2}{3} \eta_{\mu\nu} \eta_{\rho\sigma} ~~ ,
\end{eqnarray}
where $\eta_{\mu\nu}=g_{\mu\nu}-{k_{\mu}k_{\nu}}/{m_{\overrightarrow{n}}^2}$.
Here $k$ is the momentum flowing through the propagator.  Terms proportional to 
negative powers of mass of KK mode in $\eta_{\mu\nu}$ do not contribute 
as they are proportional to $k_\mu k_\nu$.  This provides a useful check on 
our calculation.  The matrix elements  have been checked for gauge invariance.  
We performed similar computation for evaluating the parton level 
subprocesses for $\gamma \gamma Z$, $\gamma ZZ$ and  $ZZZ$ productions.  
In the following we list few of the important observations.  

For the $\gamma \gamma Z$ production, in the limit 
$m_Z\rightarrow0$, we reproduce the matrix elements for $\gamma 
\gamma \gamma$ process with the changes: $(C_V^2+C_A^2)/4 
\longrightarrow Q_f^2$, $T_z \longrightarrow e$, where $C_V$, 
$C_A$ are the vector and axial vector couplings of the weak 
gauge boson respectively, $T_z = e/(\sin \theta_w \cos \theta_w)$ 
and $Q_f$ is the electric charge of the quark flavors.  In the case of
$\gamma ZZ$ production, we find that the
parton level subprocesses in SM and ADD model are similar to those 
of the $\gamma \gamma Z$ production with the changes $\gamma 
\leftrightarrow Z$.  The squared matrix elements for $\gamma ZZ$ production 
that come from 
ADD model alone are not related to those of $\gamma\gamma\gamma$ production.
The reason is that some of the terms proportional to $m_Z^2$ that appear 
in the graviton-$ZZ$ vertex cancel all the inverse power of $m_Z^2$ 
present in the $Z$ boson polarisation sum, giving contributions that 
have no analogous ones in the $\gamma\gamma \gamma$ process.  However, the 
expression for SM squared matrix elements of $\gamma ZZ$ is related to
that of $\gamma \gamma \gamma$ 
process in the SM if we take $m_Z \rightarrow 0$, 
${(C_V^4+6C_V^2C_A^2+C_A^4)}/{16}\longrightarrow Q_f^4$ and
$T_z\longrightarrow e$.
For $ZZZ$ production, squared matrix elements involving ADD vertices
do not have any relation with those of $\gamma \gamma \gamma$ production
for the same reason as $\gamma ZZ$ production case.  
The SM squared matrix elements of this process are related to those 
for the $\gamma \gamma \gamma$ process in SM 
with the following replacement
\footnote{We find that the most general result for the replacement of $n$ 
number of Z-boson with photon in the SM squared matrix element is 
$\frac{(C_V^2+C_A^2)^n + 2n(n-1)(C_V^2C_A^2(C_V^2+C_A^2)^{n-2})}{4^n}
\longrightarrow Q_f^{2n}$, which works for all the above three 
processes for $n=1,2,3$.}
in the limit $m_Z\rightarrow 0$,
 ${(C_V^6+15C_V^4C_A^2+15C_V^2C_A^4+C_A^6)}/{64}\longrightarrow 
Q_f^6$, $T_z\longrightarrow e$.
The expressions for the squared matrix elements of the processes discussed 
above are too large to be presented in this paper.
 
\section{Numerical results}
In this section, we present different kinematical distributions for the 
production of neutral triple gauge bosons.  The predictions are for the 
LHC at center of mass energy $\sqrt{S}=14$ TeV.  We have used CTEQ6L 
parton densities \cite{cteq}.  For the strong coupling constant that 
appears in CTEQ6L, we use $\Lambda_{QCD}=0.226$ GeV and $n_f=5$ flavors.  
We set the factorisation scale $\mu_F=P_T^V$ for the transverse momentum  
distribution of $V$ and $\mu_F= Q$ for the invariant mass ($Q$) distribution of 
the di-boson pair.  In addition we apply the following cuts on  $P_T^V$ 
and the rapidity $y^V$: 
\begin{eqnarray}
P_T^{\gamma,Z} \geq 25\ \mbox{GeV\qquad\qquad    and \qquad\qquad}     
y^{\gamma,Z} < 2.7 \ .
\end{eqnarray}
We also ensure that in general the invariant mass of the di-boson 
({\em i.e.}\ any two identical bosons among the 3V) is less than $M_S$.
We use $m_Z=91.1876$ GeV and $\sin^2\theta_w=0.2312$. 
The fine structure constant is taken 
as $\alpha=1/128$. 
\begin{figure}[tbh]
\centerline{
\epsfig{file=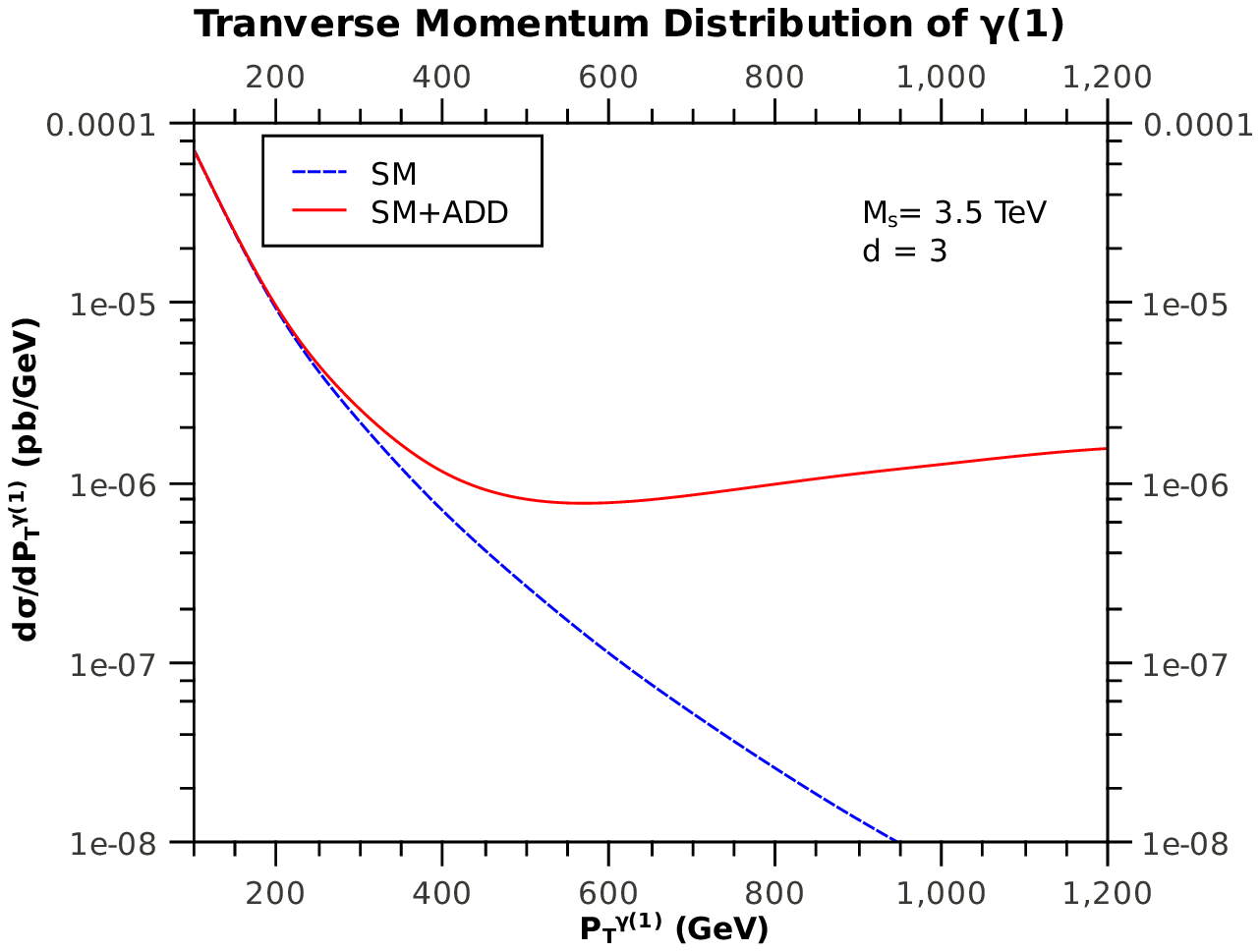,width=7cm,height=7cm,angle=0}
\epsfig{file=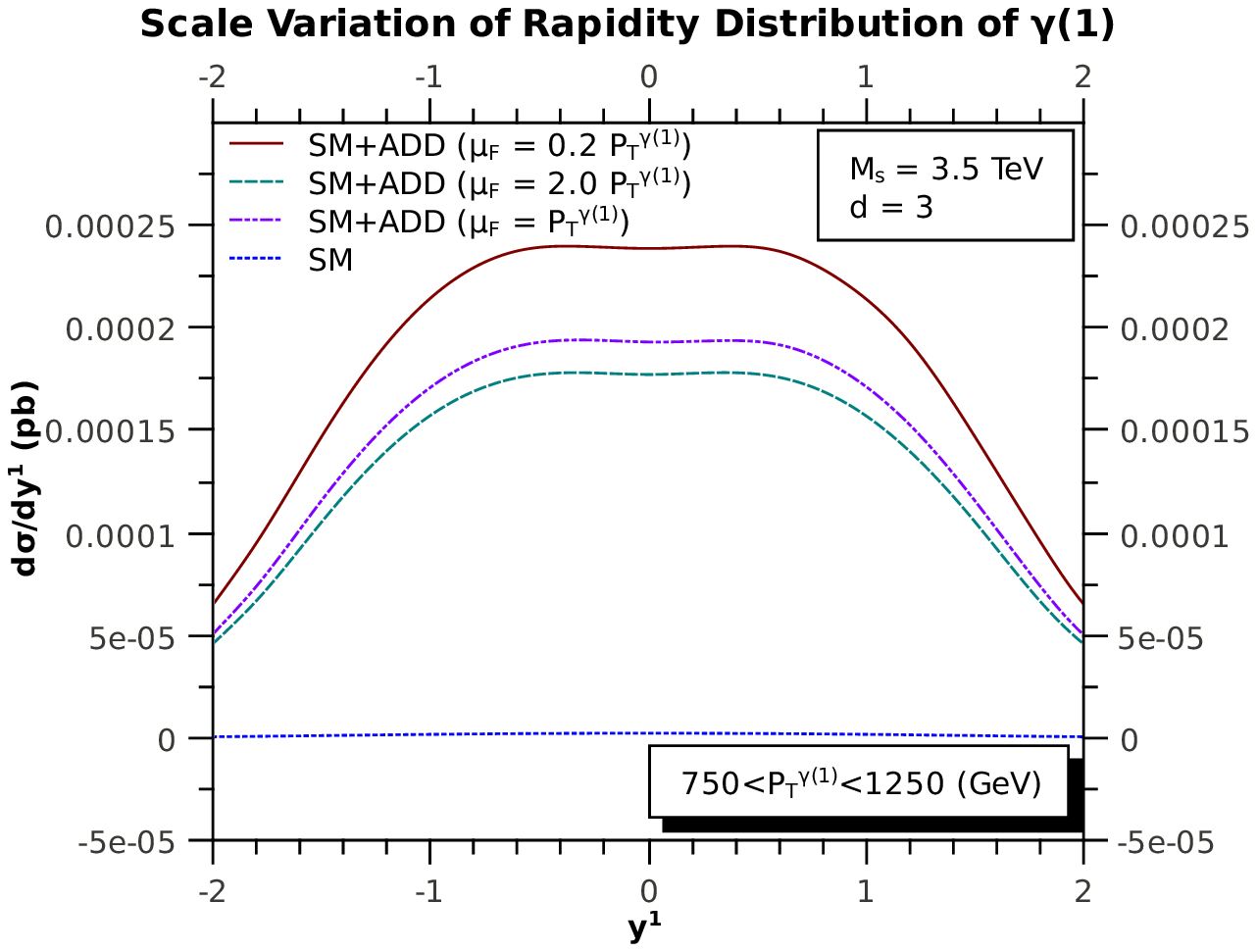,width=7cm,height=7cm,angle=0}}
\caption{Transverse momentum distribution of $\gamma_1$
[i.e, with maximum $P_T$] (left panel) at the LHC for
$M_S = 3.5$ TeV and $d = 3$.  Rapidity distribution of
$\gamma_1$ (right panel) at the LHC for ADD model parameters
$M_S = 3.5$ TeV and $d = 3$ in the region where
$P_T^{\gamma_1}\in(750,1250)$ GeV.  Dependence of rapidity
distribution of $\gamma_1$ on the factorisation scale in the
range $\mu_F=0.2 P_T^{\gamma_1}$ and $\mu_F=2 P_T^{\gamma_1}$.}
\label{3g1}
\end{figure}

Recently, both CMS \cite{CMS} and ATLAS \cite{ATLAS} reported,
searches for signatures of extra dimensions in the di-photon mass
spectrum at the LHC for 7 TeV p p collisions.  The 95 \% lower bound
on $M_S$ vary between 2.27 - 3.53 TeV depending on the number of
extra dimensions $d=7-3$ for ATLAS and $M_S$ vary between 2.3 - 3.8
TeV depending on the number of extra dimensions $d=7-2$ for CMS,
both using a fixed $K$ factor of about 1.6  \cite{diphoton1}.
Hence we have used the phenomenologically viable ADD model parameters
$M_S = 3.5$ TeV and $d=3$ for present study.

For the processes involving more than one photon, it is important to isolate 
photons from each other {\em i.e.}\ they need to be well separated in phase 
space so that they can be identified as separate objects in the detector.
To do this we consider a cone of radius $R= \sqrt{(\Delta y)^2 +
(\Delta \phi)^2}$ in the rapidity-azimuthal angle plane $(y,\phi)$ and
ensure that the minimum separation between any two photons is taken to be 
$R_{\gamma \gamma} = 0.4$.  In the following, we describe our findings for 
the various triple gauge boson production processes.
\begin{figure}[tbh]
\centerline{
\epsfig{file=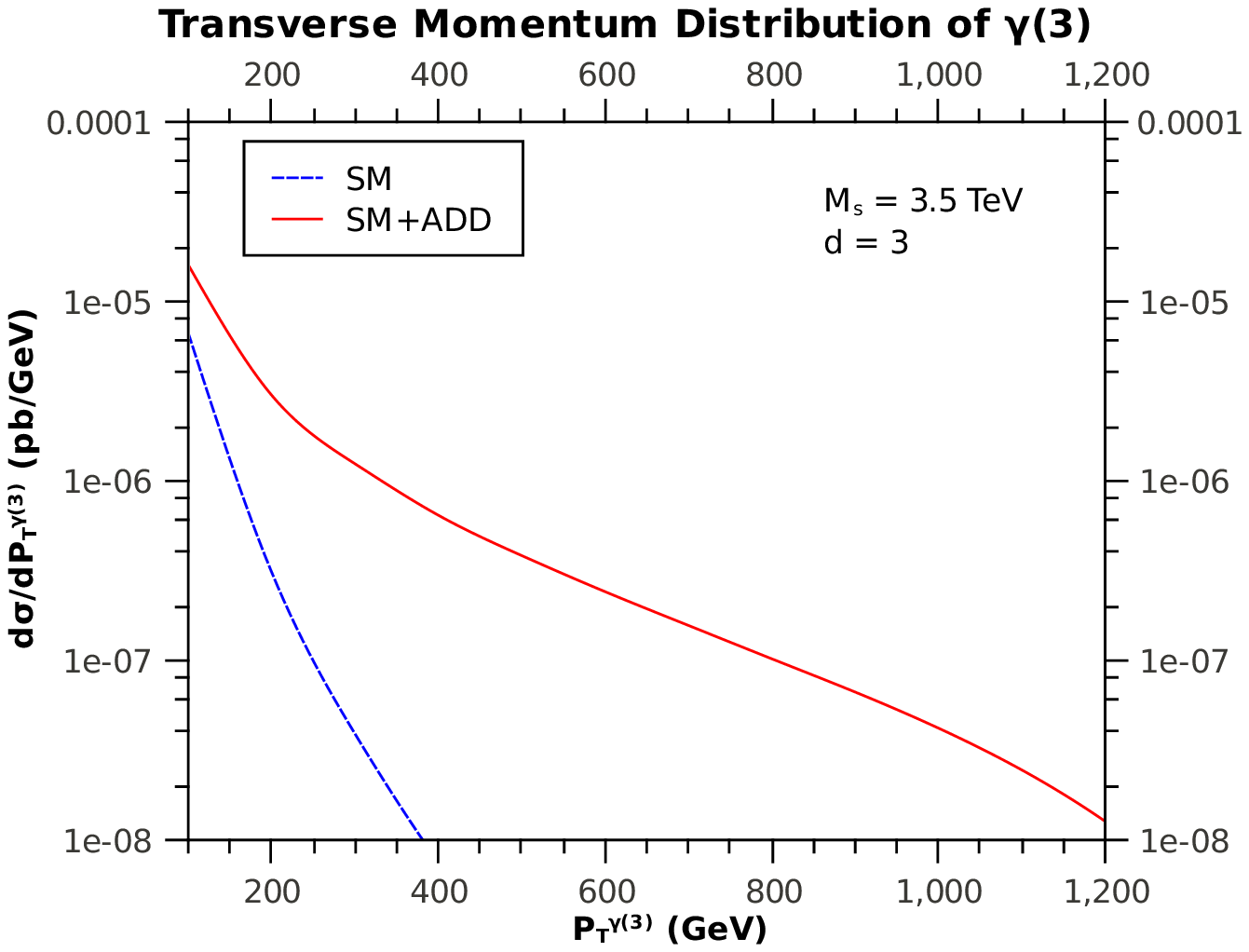,width=7cm,height=7cm,angle=0}
\epsfig{file=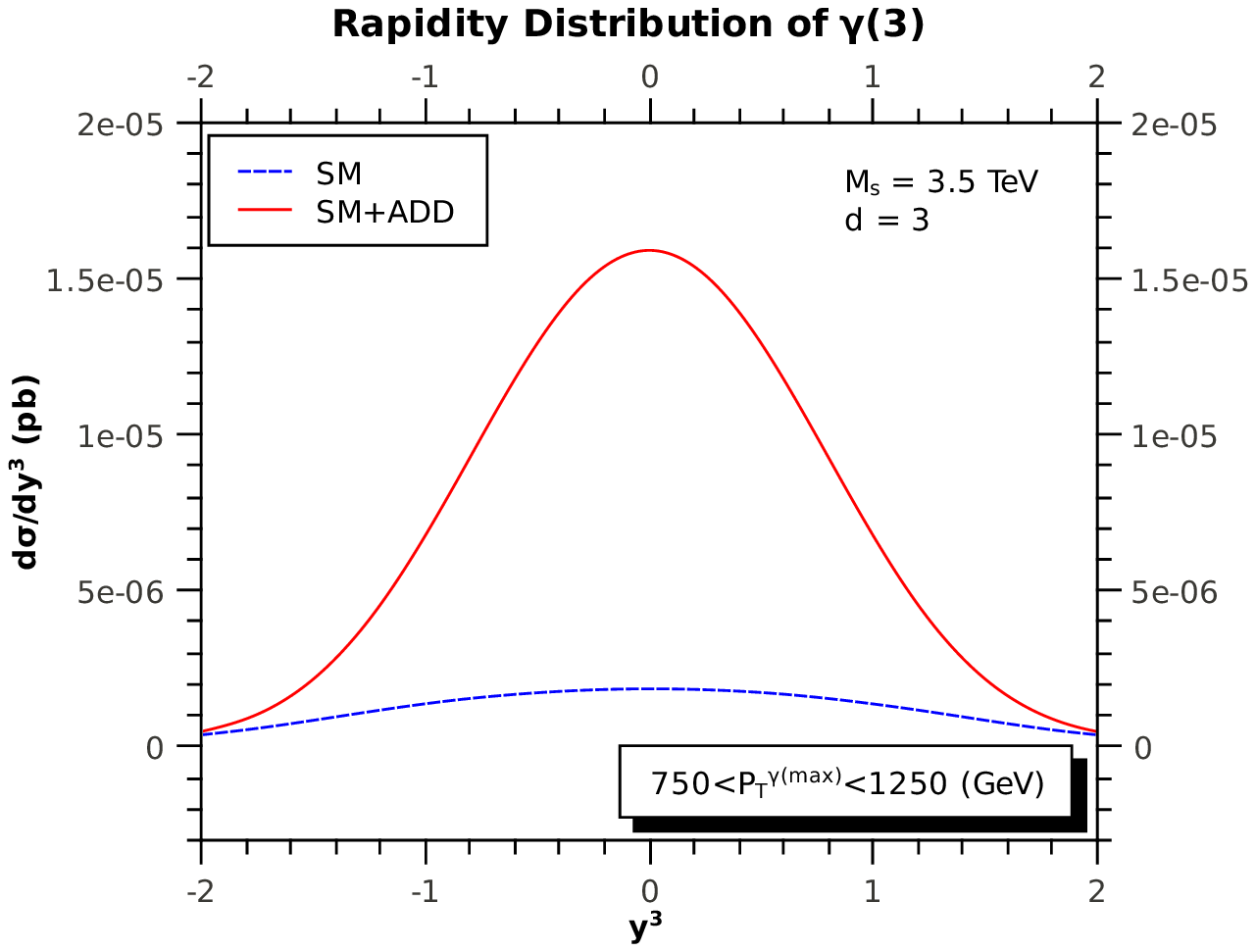,width=7cm,height=7cm,angle=0}}
\caption{Transverse momentum distribution of $\gamma_3$ 
[i.e, with minimum $P_T$] (left panel) at the LHC for $M_S = 3.5$ 
TeV and $d = 3$.
Rapidity distribution of $\gamma_3$ (right panel) at the LHC for
$M_S = 3.5$ TeV and $d = 3$ in the region where $P_T^{\gamma_3}
\in(750,1250)$ GeV.}
\label{3g3}
\end{figure}

1. $\gamma \gamma \gamma$ production:  In this case, the three photons in the
final states are 
classified in such a way that $P_T^{\gamma_1} > P_T^{\gamma_2}>
P_T^{\gamma_3}$.  We have compared our predictions for $P_T^{\gamma_1}$ 
distribution in the SM against those given in \cite{bozzi1} and found a very good 
agreement confirming the correct implementation of our analytical results 
in our numerical code.  In the left panel of Fig.\ \ref{3g1}, we present 
the transverse momentum distributions of $\gamma_1$ in SM as well as in SM+ADD. 
We have chosen $M_S=3.5$ TeV and $d=3$ as representative parameters of the ADD model.  
In high $P_T^{\gamma_1}$ region, the distribution for SM+ADD is fully
controlled by processes coming from ADD model and is enhanced due to the   
dominant contributions of the KK modes.  In the right panel of Fig.\ \ref{3g1}, rapidity
distributions of the most energetic photon $\gamma_1$ are shown for $750 < 
P_T^{\gamma_1} < 1250$ GeV in SM and SM+ADD.  It is seen that the SM 
contribution is extremely small in this range.  

In order to estimate the factorisation scale $\mu_F$ dependence 
present in our LO results, in the right panel of Fig.\ \ref{3g1} 
we have plotted rapidity distributions for three different 
choices of $\mu_F$ {\em i.e.}, $\mu_F=(0.2,1,2) P_T^{\gamma_1}$.  
In the central rapidity region the variation of the rapidity distribution 
with respect to the factorisation scale is the largest.  
With respect to the central choice of $\mu_F=P_T^{\gamma_1}$, 
the variation is about $23.6 ~\%$ and $8.2 ~\%$ for the choice of 
$\mu_F=0.2~ P_T^{\gamma_1}$ and $\mu_F =2~ P_T^{\gamma_1}$ respectively.

The $P_T$ distribution of $\gamma_2$ is found to be similar to that 
of $\gamma_1$, but is different for $\gamma_3$ (the least 
energetic of the three photons) as shown in Fig.\ \ref{3g3} 
(left panel).  Similarly its rapidity distribution Fig.\ \ref{3g3} 
(right panel) is also different from the most energetic photon.
\begin{figure}[tbh]
\centerline{
\epsfig{file=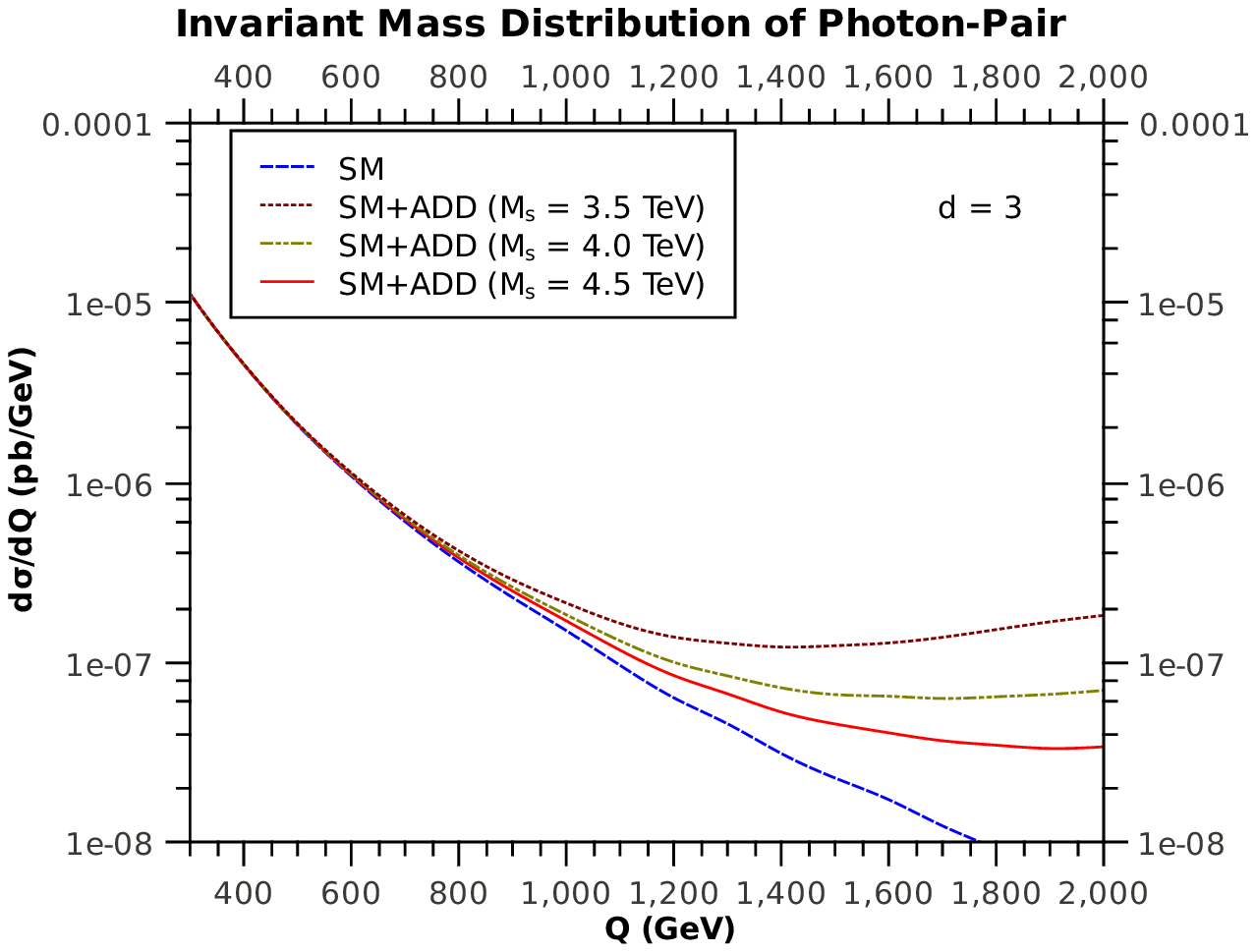,width=6cm,height=6cm,angle=0}
\epsfig{file=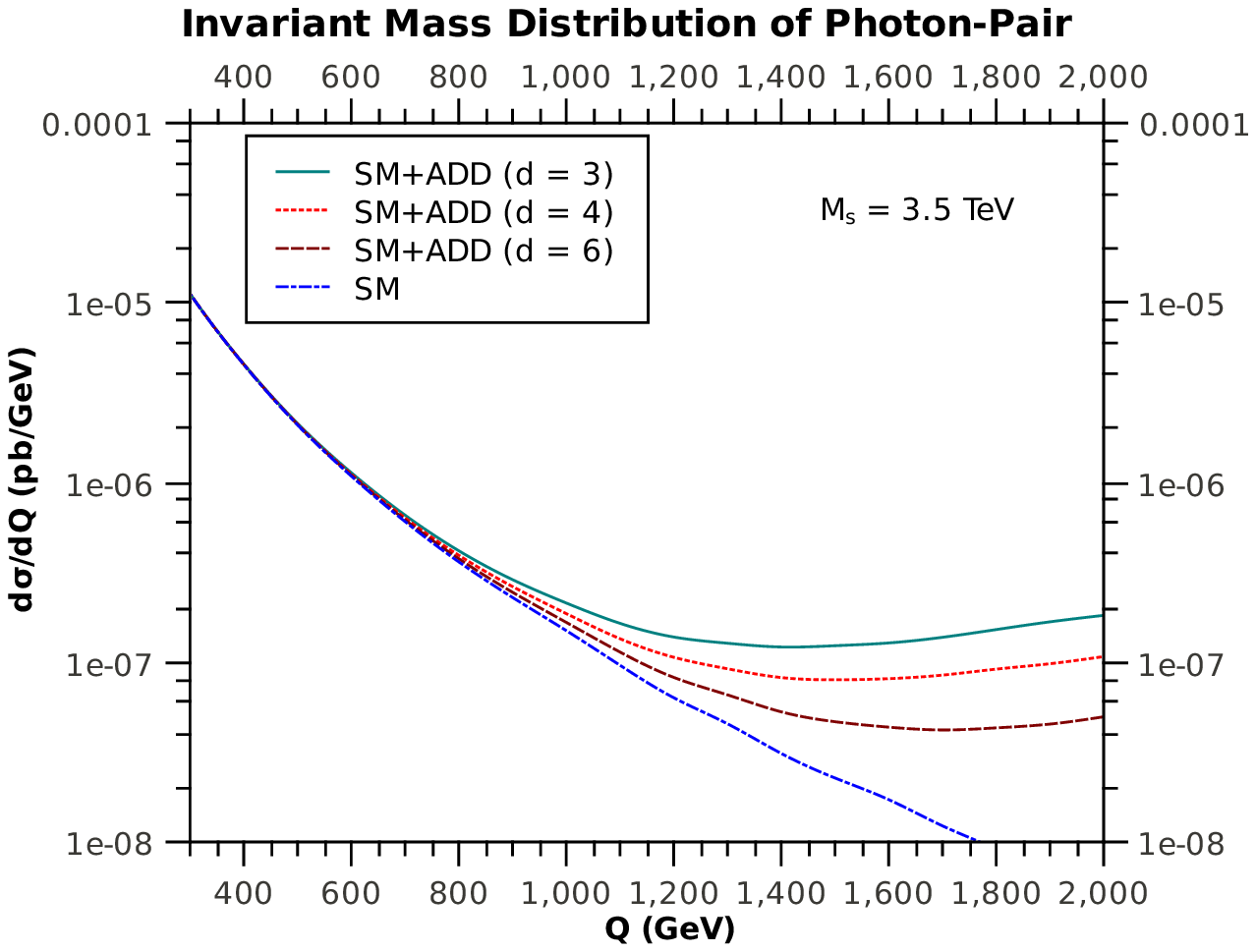,width=6cm,height=6cm,angle=0}}
\centerline{
\epsfig{file=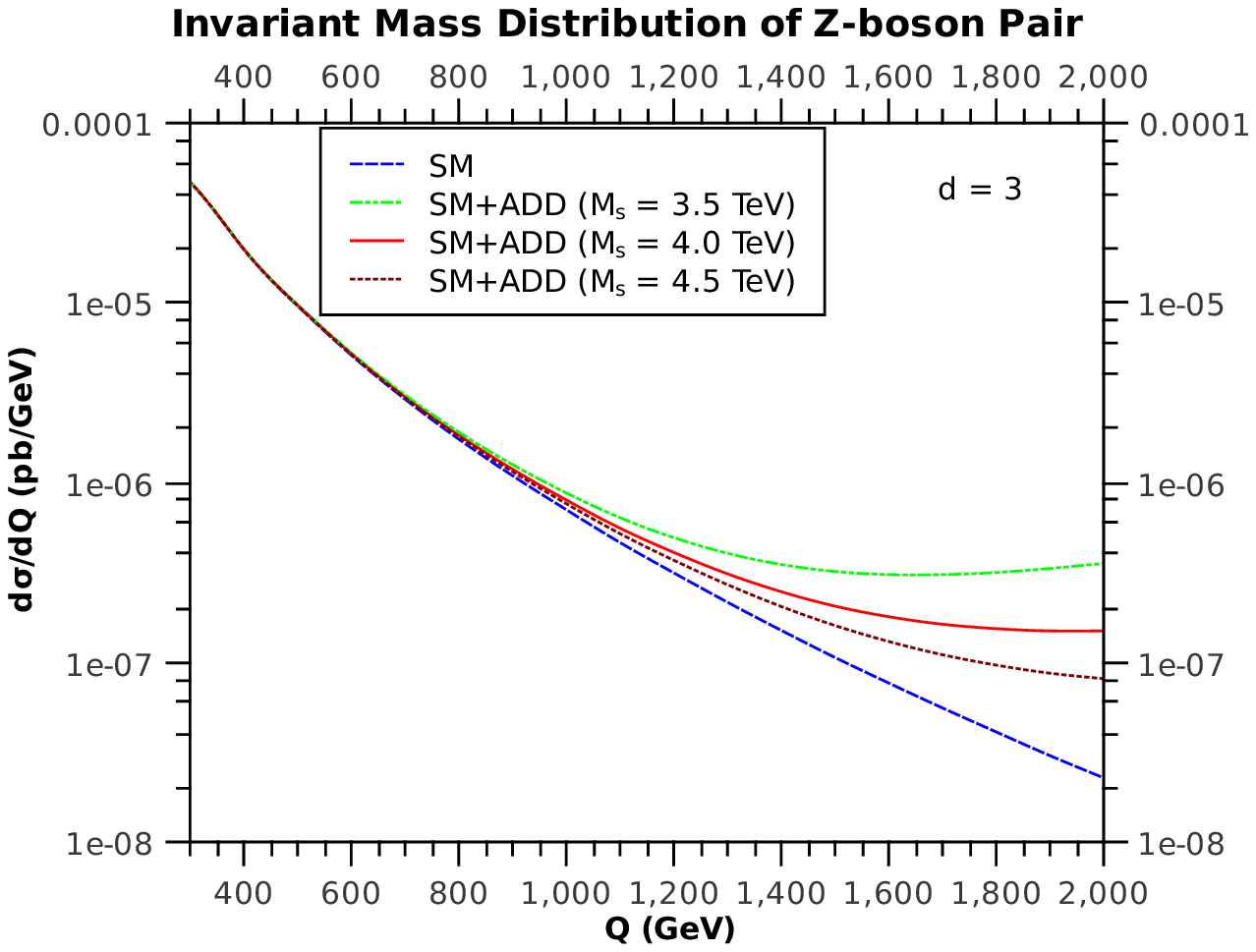,width=6cm,height=6cm,angle=0}
\epsfig{file=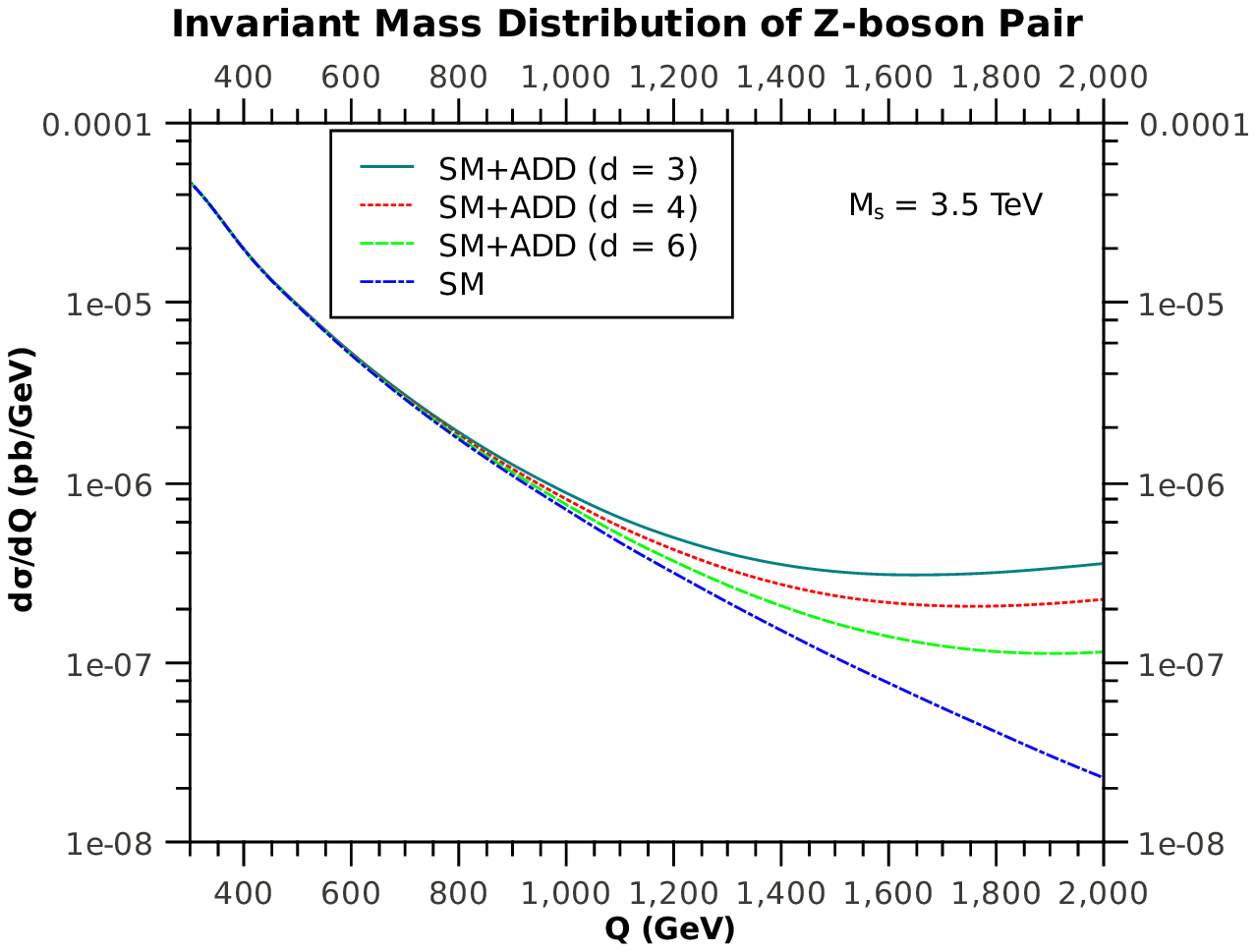,width=6cm,height=6cm,angle=0}
}
\caption{Invariant mass distribution of the photon pair (top left panel)
and Z boson pair (bottom left panel) at the LHC for $d = 3$ with different values 
of $M_S$.  For $M_S = 3.5$ TeV with different values of $d$,
the invariant mass distribution of the photon pair (right top panel) and 
Z pair (right bottom panel).}
\label{inv2g}
\end{figure}

\begin{figure}[tbh]
\centerline{
\epsfig{file=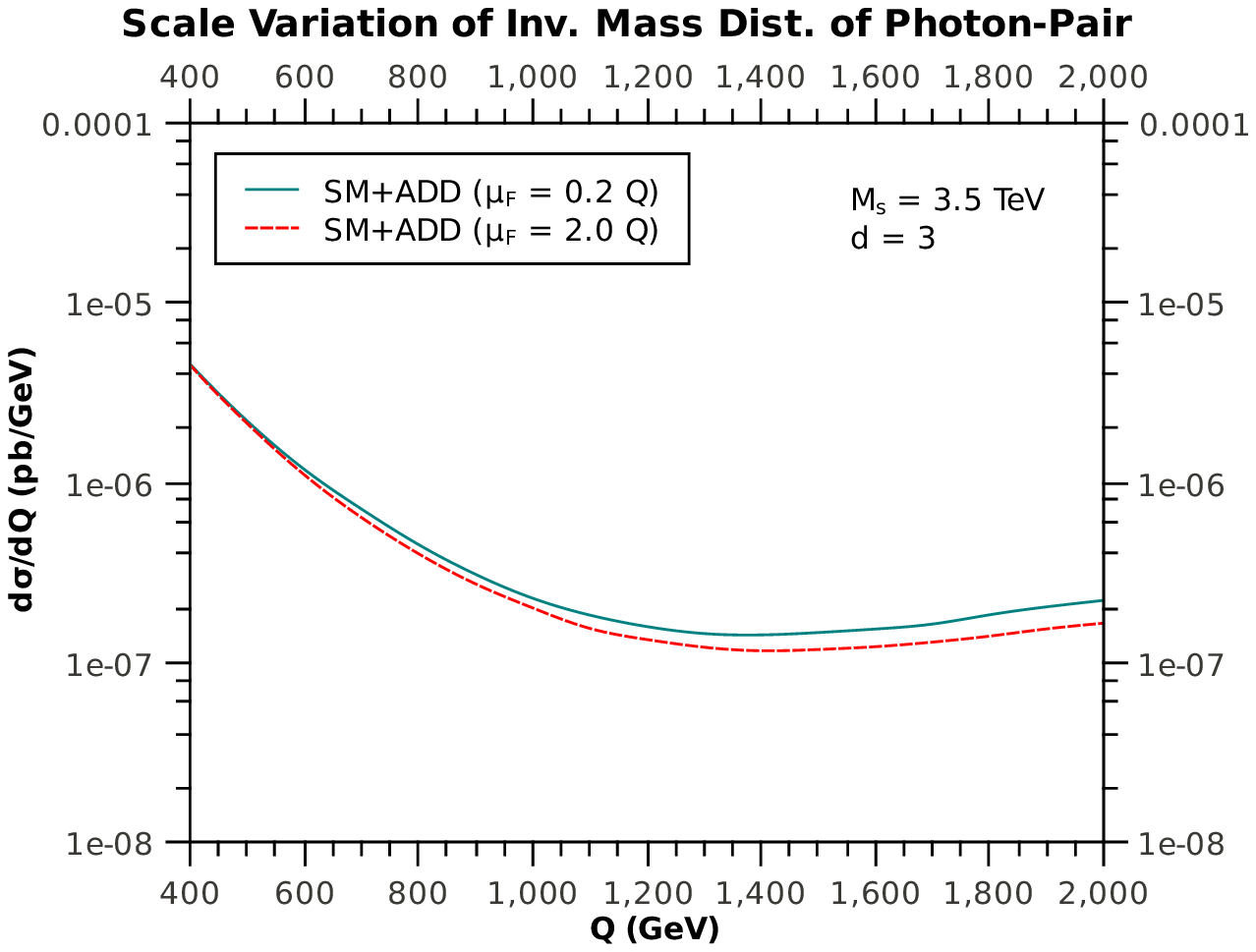,width=6cm,height=6cm,angle=0}
\epsfig{file=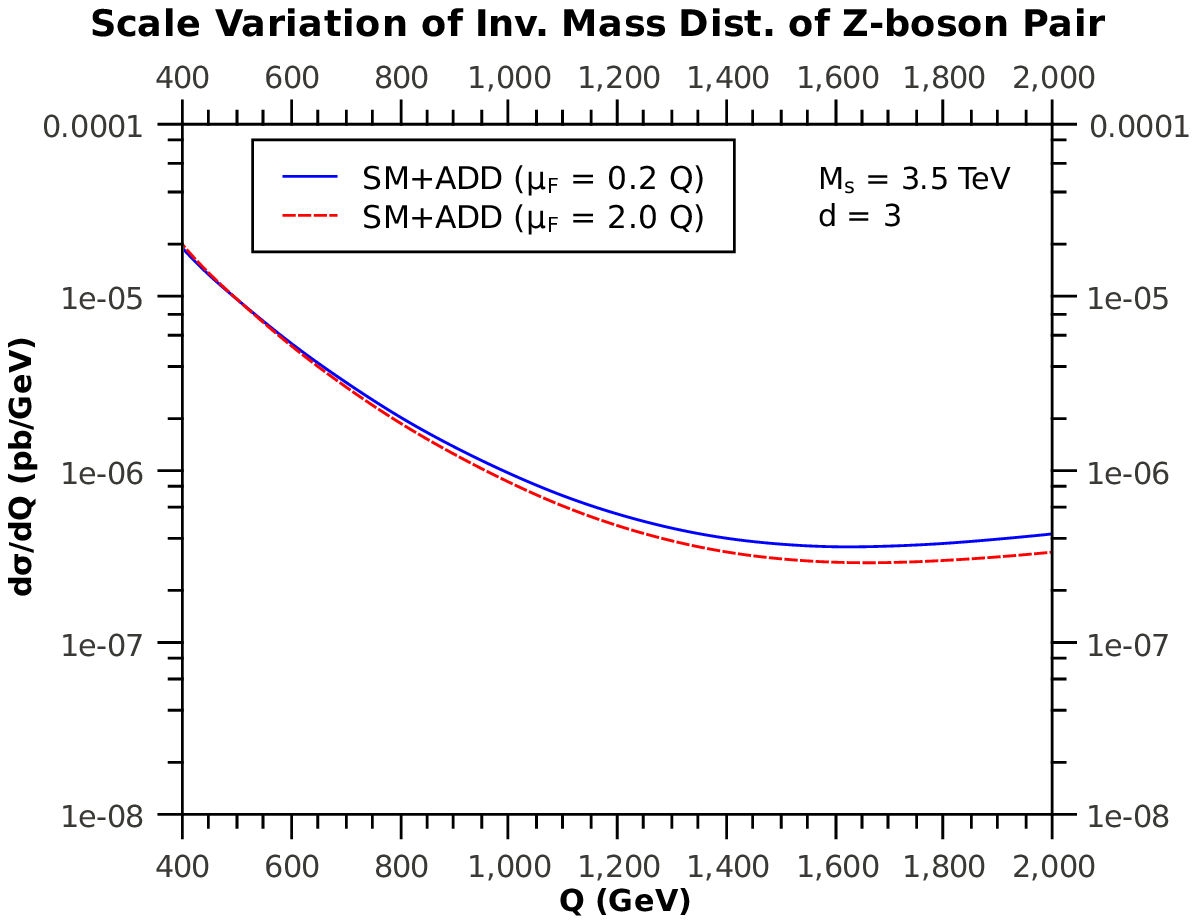,width=6cm,height=6cm,angle=0}}
\caption{Dependence of invariant mass distribution of the photon pair
(left panel) on factorisation scale at the LHC for $d=3$ and $M_S=3.5$ 
TeV.  Dependence of invariant mass distribution of Z boson pair (right
panel) on factorisation scale at the LHC for $d = 3$ and $M_S = 3.5$ 
TeV.}  
\label{scale2z}
\end{figure}

2. $\gamma \gamma Z$ production:  Here, the invariant 
mass distribution of the photon pair is a useful observable because in the 
ADD model the photon pair is one of the clean decay modes of the KK 
graviton and in the region of interest, this could give an enhancement of 
the tail of the distribution.  In Fig.\ \ref{inv2g} (top left panel) we 
have presented the invariant mass distributions of the photon pair for 
different choices of $M_S=(3.5,4,4.5)$ TeV fixing $d=3$, while in the top
right panel the same distribution is plotted for different choices of $d=3,4,6$ 
but a fixed $M_S=3.5$ TeV.  We find that the KK modes dominate over the SM 
contribution for larger values of invariant masses (say above $400$ GeV 
for a given set of $M_S$ and $d$ values) of photon pairs 
leading to a significant enhancement of the signal over the background.
We plot the factorisation scale dependence of invariant mass
distributions of photon pairs in Fig.\ \ref{scale2z} (left panel) for
different choices of $\mu_F$, {\em i.e.}, $\mu_F=(0.2,2) Q$.


3. $\gamma Z Z$ production:  Invariant mass of Z boson pair  is again a
useful observable.   We have done a similar analysis as we did for 
$\gamma \gamma Z$ and use the same choice of factorisation scale and ADD 
model parameters.  The invariant mass distributions are shown in the lower 
panels of Fig.\ \ref{inv2g} for different choices of  $M_S$ and $d$.  We 
find that the invariant mass distributions of  photon pairs in $\gamma 
\gamma Z$ production and Z boson pairs here have similar qualitative 
behavior.
In order to investigate the uncertainty resulting from the factorisation 
scale $\mu_F$, in Fig.\ \ref{scale2z} (right panel), we have plotted the 
invariant mass distributions of Z pairs for different choices of $\mu_F$, 
{\em i.e.}, $\mu_F=(0.2,2) Q$.  

\begin{figure}[tbh]
\centerline{
\epsfig{file=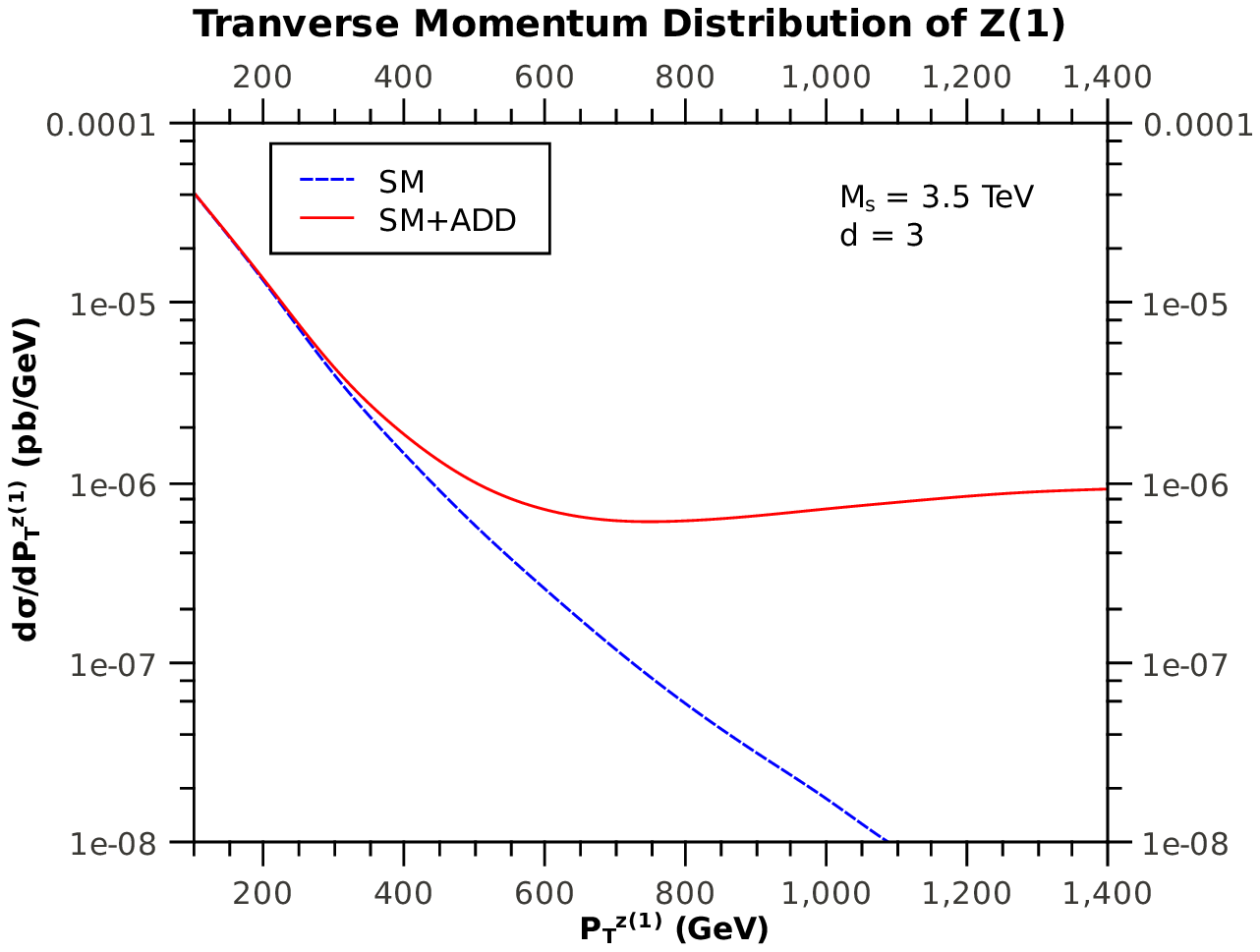,width=5cm,height=6cm,angle=0}
\epsfig{file=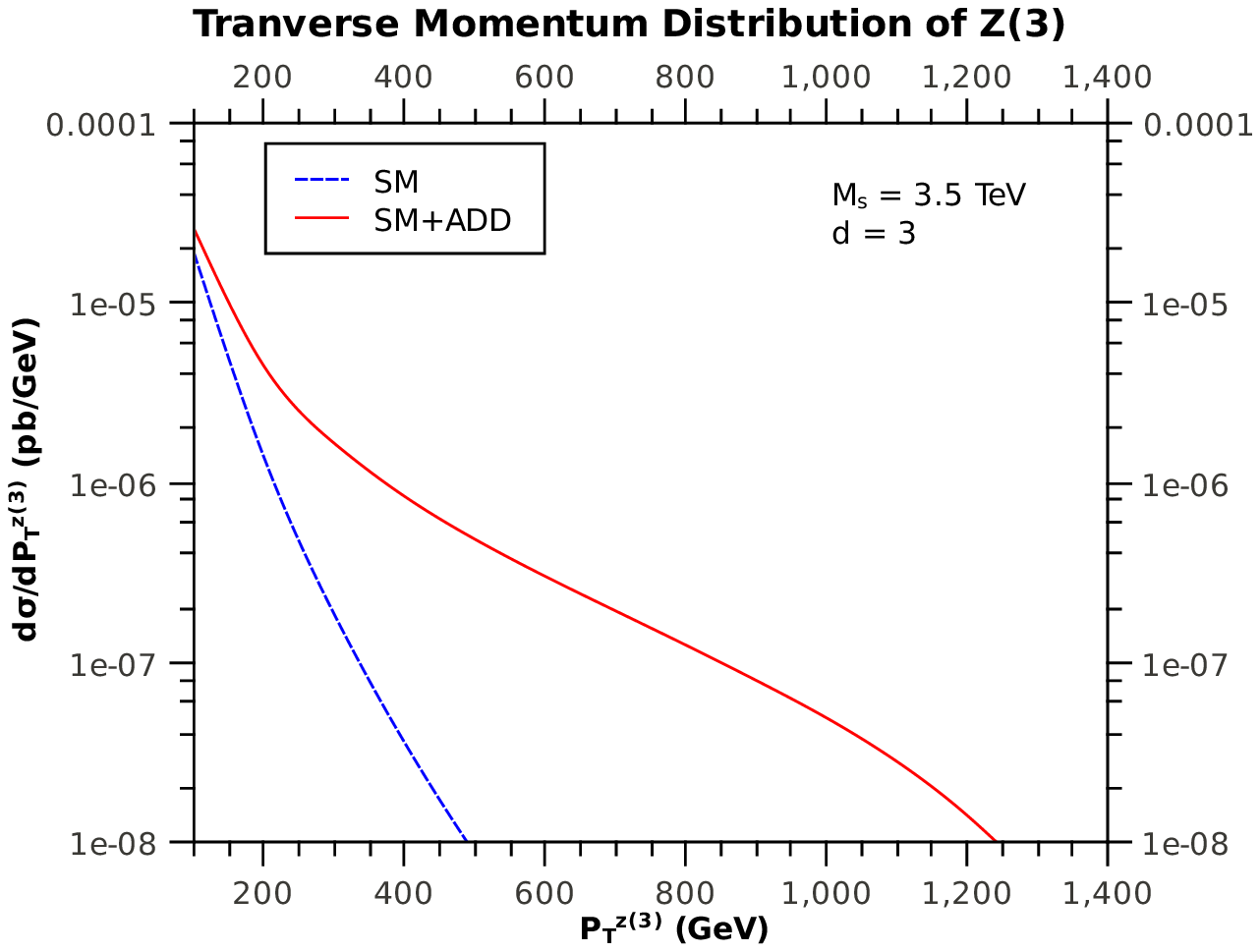,width=5cm,height=6cm,angle=0}
\epsfig{file=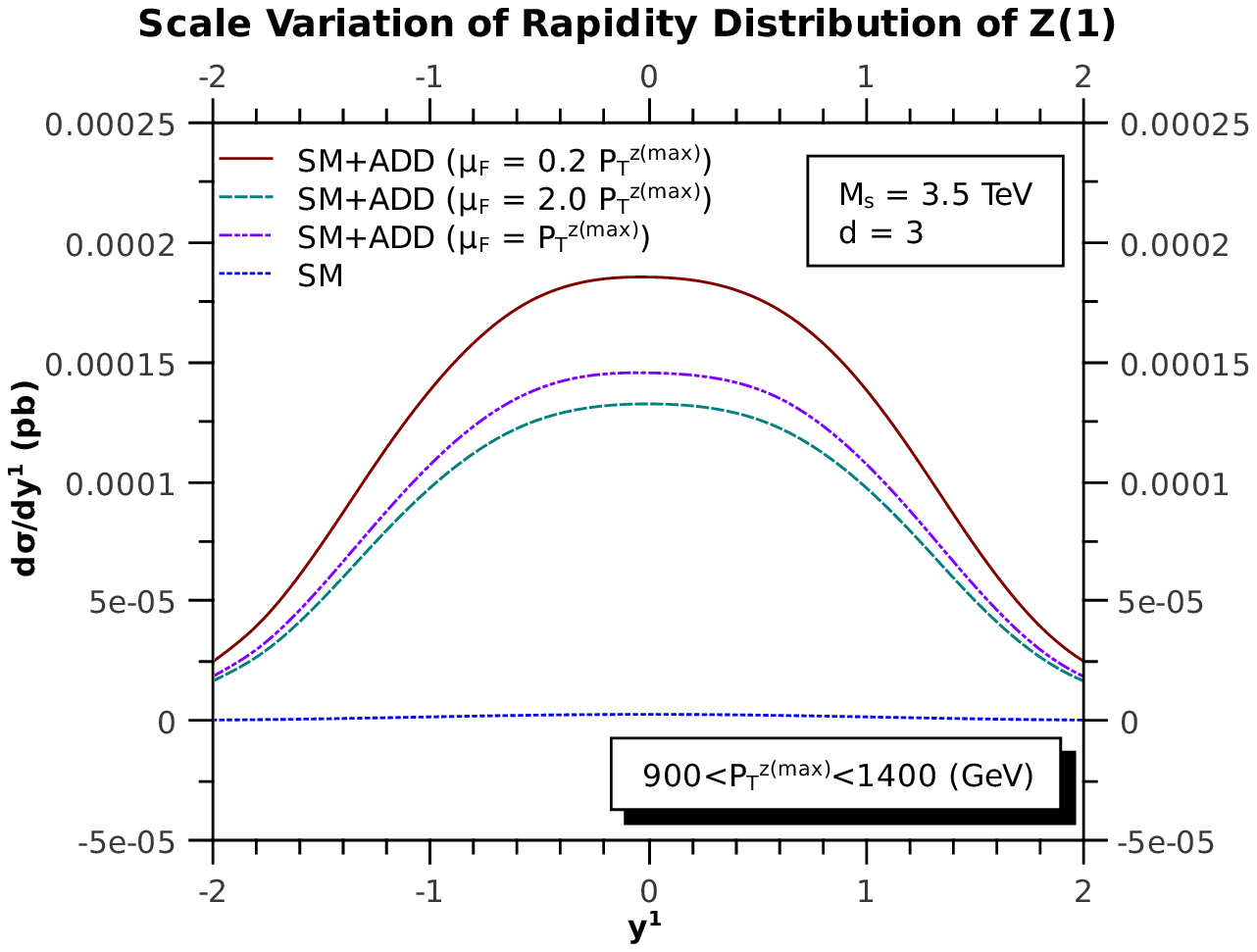,width=5cm,height=6cm,angle=0}}
\caption{Transverse momentum distribution of $Z_1$ boson [i.e, with maximum 
$P_T$] (left) and $Z_3$ boson [i.e, with least $P_T$] (middle) at the LHC 
for $M_S = 3.5$ TeV and $d = 3$.  Rapidity distribution of that $Z_1$ boson 
(right) at the LHC for $M_S = 3.5$ TeV and $d = 3$ in the region where 
$P_T^{Z_1}\in(900,1400)$ GeV.}
\label{3z1}
\end{figure}

4. $ZZZ$ production:  We have classified the triple $Z$-bosons in such 
a way that $P_T^{Z_1}>P_T^{Z_2}>P_T^{Z_3}$ and for the $P_T^{Z_i}$ 
distribution we make the choice of factorisation scale $\mu_F=P_{T}^{Z_i}$.
In Fig.\ \ref{3z1}, we have presented the transverse momentum distributions 
of $Z_1$ (left panel) and $Z_3$ (middle panel) and rapidity distribution 
of $Z_1$ (right panel) for SM and SM+ADD with $M_S=3.5$ TeV and $d=3$.  
For  the rapidity distribution, we 
have constrained $900 < P_T^{Z_1} <1400$ GeV.  
As in the case of $\gamma \gamma \gamma$, the $P_T^{Z_2}$ distribution is 
similar to that of $P_T^{Z_1}$ distribution.  
We have also shown the 
sensitivity of rapidity distribution to the factorisation scale 
$\mu_F$ by varying between $\mu_F=0.2 P_T^{Z_1}$ and $\mu_F=2 P_T^{Z_1}$.  
In the central rapidity region we estimate the variation of the rapidity 
distribution with the factorisation scale and find that for $\mu_F=0.2 
P_T^{Z_1}$ and $\mu_F=2 P_T^{Z_1}$, the variation is about 
$27.5 ~\%$ and $8.9 ~\%$ respectively with respect to $\mu_F=P_T^{Z_1}$.
The rapidity distribution for $Z_2$ is similar to that of $Z_1$ while 
$Z_3$ is different.

\begin{figure}[tbh]
\centerline{
\epsfig{file=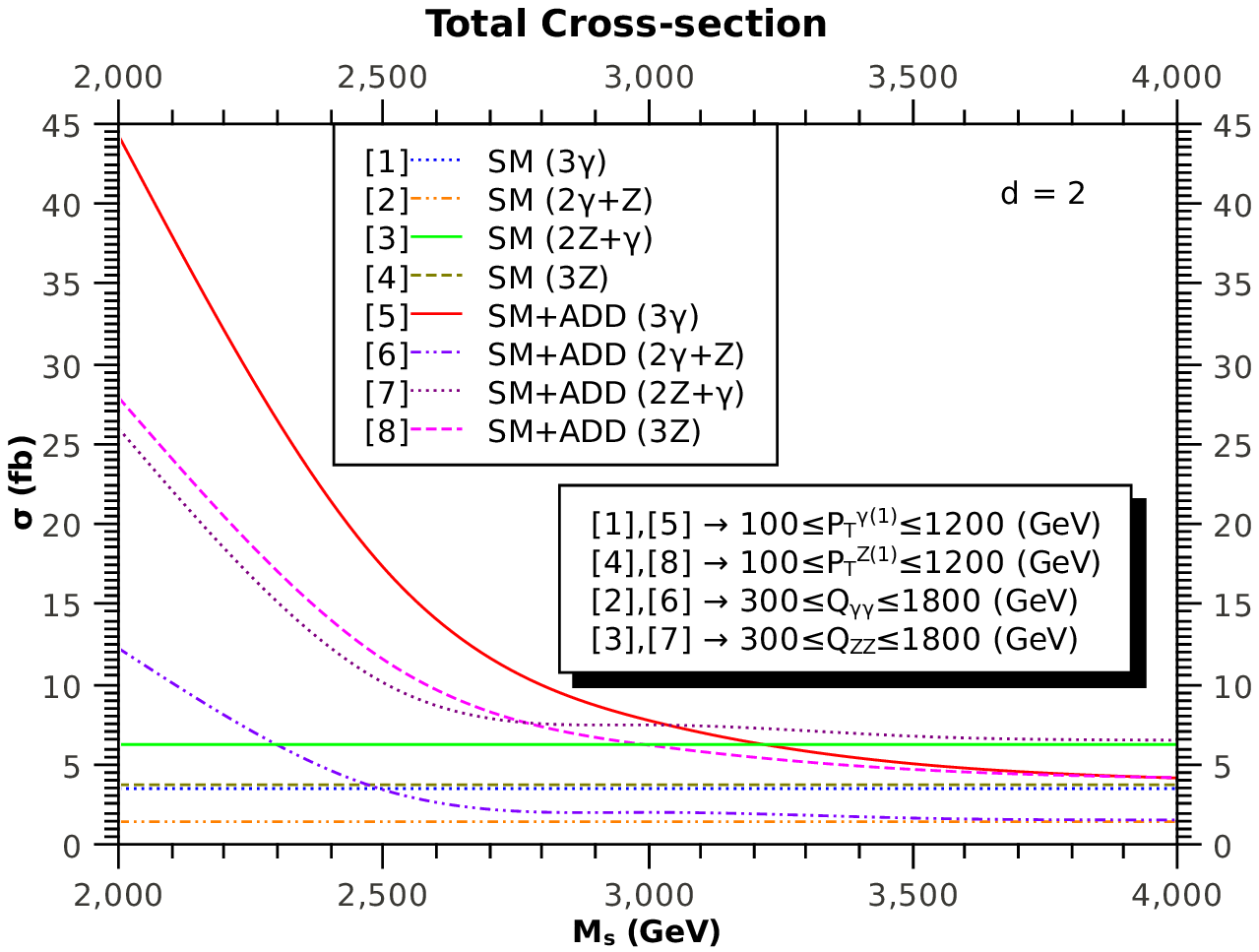,width=10cm,height=9cm,angle=0}
}
\caption{The total cross section for all the triple neutral gauge boson 
as a function of the extra dimension scale $M_S$ for number of extra 
dimensions $d=2$.  The horizontal lines corresponds to the various 
SM contribution.
}
\label{tcs}
\end{figure}

5. Total cross section: The total cross sections for various processes
involving neutral triple gauge boson final states as a function of $M_S$
for a fixed value of $d=2$ are given in Fig.\ \ref{tcs}.  The SM 
contributions that do not depend on ADD model parameter $M_S$ appear as
horizontal lines.

\begin{figure}[tbh]
\centerline{
\epsfig{file=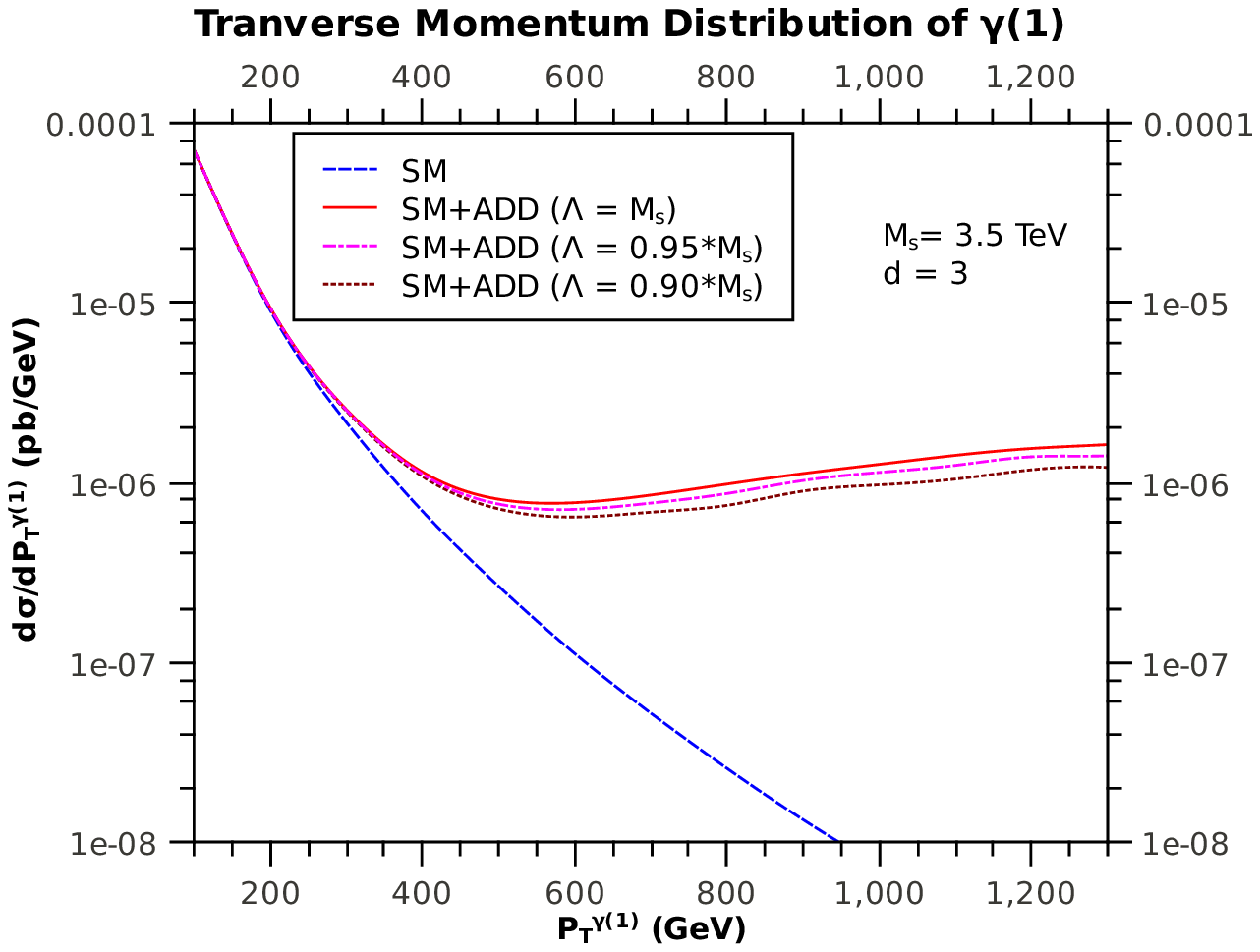,width=6cm,height=6cm,angle=0}
\epsfig{file=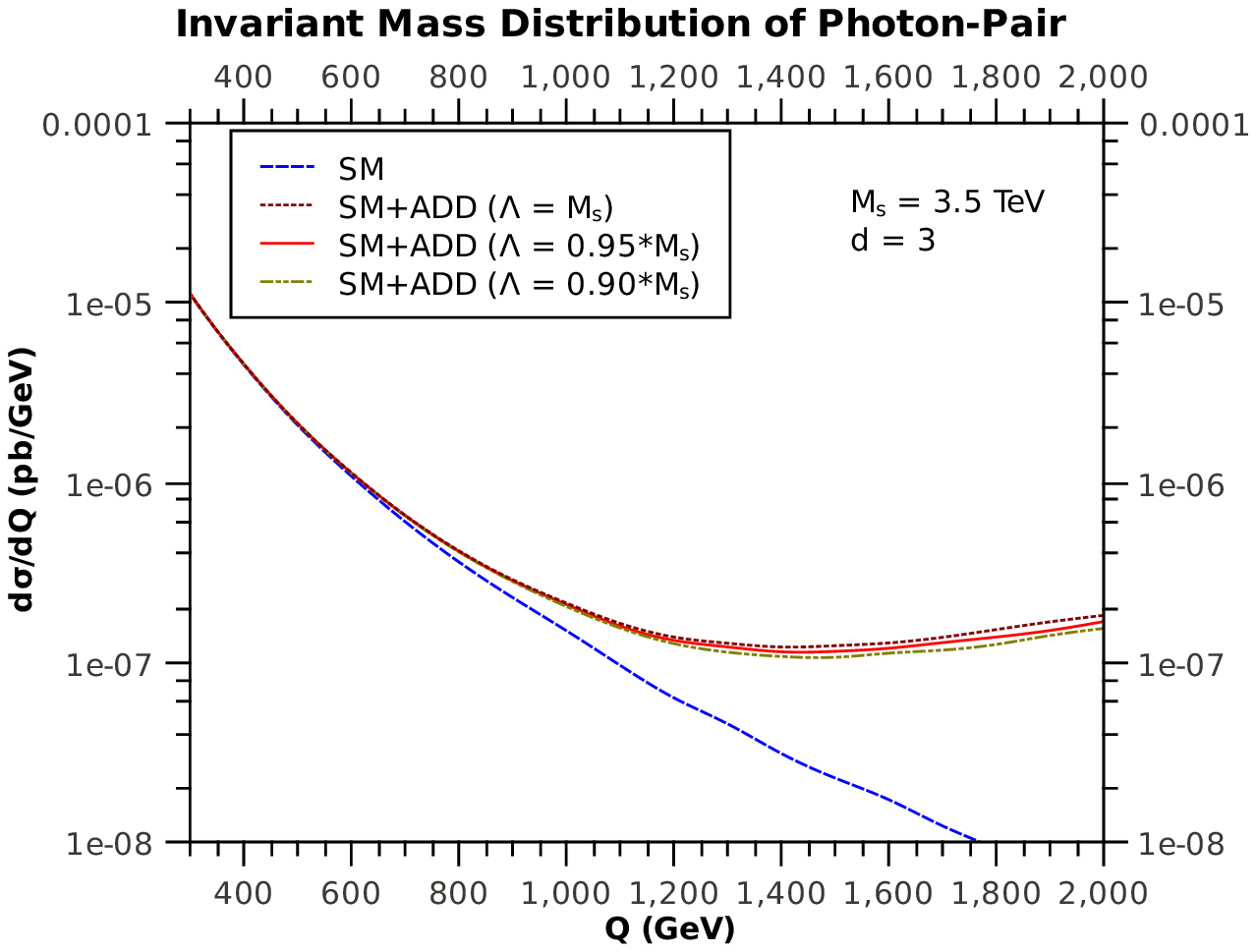,width=6cm,height=6cm,angle=0}}
\caption{$P_T^{\gamma_1}$ distribution of the $\gamma \gamma
\gamma$ final state (left) and invariant mass distribution of 
the $\gamma \gamma$ pair in the $\gamma \gamma Z$ final state 
(right), for the cut off scale $\Lambda=(0.9,~ 0.95,~ 1) M_S$, 
where $M_S=3.5$ TeV and $d=3$.} 
\label{Lambda}
\end{figure}

So far in our numerical analysis we have put the UV cutoff 
$\Lambda = M_S$ which is the conventional choice to do the 
phenomenology.  The sensitivity of the choice of UV cutoff
is presented in Fig.\ \ref{Lambda} for $P_T^{\gamma_1}$ distribution 
of $\gamma \gamma \gamma$ final state and also the invariant 
mass distribution of $\gamma \gamma$ pair of $\gamma \gamma Z$ 
process by varying $\Lambda = (0.9,~ 0.95,~ 1) M_S$.
The cross section at $P_T^{\gamma_1}=1200$ GeV 
varies between 10 - 24 \% as we vary $\Lambda = (0.9,~ 0.95) M_S$
as compared to $\Lambda = M_S$ for the $\gamma \gamma \gamma$ 
process.  Similarly the cross section for $\gamma \gamma Z$
process at $Q=2000$ GeV the variation is 7 - 15 \% in the same 
rage of $\Lambda$.

The SM integrated K-factor for the $\gamma \gamma \gamma$ process is about 
2.6 and for $\gamma \gamma Z$ it is 1.5 and these are larger than 
uncertainties resulting from the factorization scale
variations at LO \cite{bozzi1}.  This warrants 
a full next to leading order QCD analysis which is beyond the scope of this 
paper.  Such an analysis is reserved for future publication \cite{tobe1}.  
  
\section{Conclusion}

In this paper, we have studied the neutral triple gauge boson production 
at the LHC in theories with large extra dimensions which are 
produced via the exchange of a tower of KK graviton taking into account 
the SM contributions.  All the final state photons and Z-bosons are taken to be 
real.  We have performed various checks on our analytical results, and
the numerical predictions are obtained using a Monte Carlo code which 
allows us to implement various experimental cuts.  Our code has also 
been tested to reproduce SM transverse momentum distribution of the 
largest transverse momentum for the $\gamma \gamma \gamma$ process.  
For the case in which the gauge bosons in the final state are 
identical we have presented the transverse momentum distribution by 
ordering the transverse momentum as $P_T^{1} > P_T^{2} > P_T^{3}$.  
We find that $P_T^1$ and $P_T^2$ 
distributions are similar but the one for $P_T^3$ is different.  
The rapidity distributions are also presented. 
For the case where one of the gauge bosons in the final 
state is different we choose to use the invariant mass distribution of 
the identical di-bosons as it would be a better discriminator in the 
region of interest.  We have also studied the dependency of the ADD model 
parameter $M_S$ and the number of extra dimensions $d$ keeping
the UV scale $\Lambda=M_S$.  In addition we have reported the 
sensitivity of the choice of $\Lambda$ by varying it from $\Lambda = 0.9 M_S$ 
to $0.95 M_S$.  We have also studied the dependence of our LO predictions on 
the factorisation scale. 

\section*{Acknowledgments}
The work of M.C.K.\ has been supported by Deutsche Forschungsgemeinschaft in
Sonderforschungsbereich/Transregio 9 and by the European Commission through
contract PITN-GA-2010-264564 (LHCPhenoNet).
The work of V.R.\ has been partially supported by funds made available to 
the Regional Centre for Accelerator based Particle Physics (RECAPP) by 
the Department of Atomic Energy, Govt. of India.  
S.S.\ would like to thank UGC, New Delhi for financial support.  

\eject

\newpage

\end{document}